\documentclass[aps,prd,twocolumn,floatfix,noshowpacs,tightenlines,noshowkeys,superscriptaddress,amsmath,amssymb,nofootinbib]{revtex4-1}
\usepackage{amssymb,amsbsy,epsfig,color,graphicx}
\usepackage{color}
\usepackage{[longtable}
\usepackage{array}
\usepackage{dcolumn}   % needed for some tables
\usepackage{cellspace}
\usepackage{mathtools}
\usepackage{amstext}
\usepackage{amssymb}
\usepackage{stmaryrd}
\usepackage{stackrel}
\usepackage{graphicx}
\usepackage{esint}
\usepackage[utf8]{inputenc}
\usepackage{adjustbox}
\usepackage{multirow}
\usepackage{float}
\usepackage{booktabs}
\usepackage{enumitem}
\usepackage{natbib}
\usepackage{slashed}
\usepackage[hidelinks]{hyperref}
\hypersetup{
  colorlinks   = true, %Colours links instead of ugly boxes
  urlcolor     = blue, %Colour for external hyperlinks
  linkcolor    = blue, %Colour of internal links
  citecolor   = red %Colour of citations
}
\usepackage{etoolbox} % for \appto
\usepackage{lipsum} % for mock text
\usepackage[capitalize]{cleveref}
\usepackage{lipsum,booktabs}
\usepackage{changepage}
\usepackage{multirow}
\usepackage[caption=false]{subfig}
\allowdisplaybreaks
\usepackage{soul}
\usepackage{ragged2e}
\usepackage{bm}
\begin{document}

\preprint{APS/123-QED}

\title{Activity Induced Diffusion Recovery in Crowded Colloidal Suspension}

% Force line breaks with \\
%\thanks{Author to whom correspondence should be addressed: k.dey@iitgn.ac.in}%

\author{Arnab Maiti}
%\email{maiti_arnab@iitgn.ac.in}
\affiliation{Laboratory of Soft and Living Materials, Discipline of Physics, Indian Institute of Technology Gandhinagar, Palaj, Gujarat 382055, India}

\author{Yuki Koyano}
\affiliation{Graduate School of Human Development and Environment, Kobe University, 3-11 Tsurukabuto, Nada-ku, Kobe, Hyogo 657-0011, Japan}


\author{Hiroyuki Kitahata}
\affiliation{Department of Physics, Graduate School of Science, Chiba University, Yayoi-cho 1-33, Inage-ku, Chiba 263-8522, Japan}

\author{Krishna Kanti Dey}
\email{Author to whom correspondence should be addressed: k.dey@iitgn.ac.in}
\affiliation{Laboratory of Soft and Living Materials, Discipline of Physics, Indian Institute of Technology Gandhinagar, Palaj, Gujarat 382055, India}
\date{\today}% It is always \today, today,
             
\begin{abstract}
We show that the force generated by active enzyme molecules are strong enough to influence the dynamics of their surroundings under artificial crowded environments. We measured the behavior of polymer microparticles in a quasi-two-dimensional system under aqueous environment, at various area fraction values of particles. In the presence of enzymatic activity not only the diffusion of the suspended particles at shorter time-scale regime enhanced, the system also showed a transition from sub-diffusive to diffusive dynamics at longer time-scale limits. Similar observations were also recorded with enzyme functionalized microparticles. Brownian dynamics simulations have been performed to support the experimental observations.  
\end{abstract}
\maketitle
%\section{\label{sec:level1}INTRODUCTION}\protect\
Cellular functions usually involve many enzymes mediated catalytic reactions that control the rate of various chemical transformations~\cite{1}. Unlike the biomolecular motors, most enzymes in nature operate in free states and rather diffuse to and away from their substrates during catalytic reactions~\cite{2}. For a long time, such molecules were believed neither to have any energy transduction ability during substrate turnover nor to influence significantly the dynamics of their surroundings. In a series of recent experiments, enzymes have, however, been found to generate forces during substrate turnover, which were significant enough to influence their dynamics and that of their surroundings in aqueous solutions~\cite{3,4, 5, 6, 7, 8, 9}. The behavior is analogous to the generation of randomly fluctuating forces inside cells by the aggregation and adaptations of molecular motors, which are believed to drive diffusive-like, non-thermal motion of cellular components, effecting the overall metabolic state of the cell~\cite{10, 11, 12}. Population of active bacteria has also been observed to considerably affect the dynamics of their surroundings either by direct interactions (hydrodynamic coupling)~\cite{13}, collisions~\cite{14}, or by changing the fluid rheology considerably~\cite{15, 16, 17, 18}. These observations provide strong motivation to investigate if enzymes, while catalyzing various chemical reactions within intracellular crowded environments, could generate sufficient mechanical forces to influence the motion of nearby particles. A positive answer to this will refine our understanding of organelle and small molecules' motion in cells and underscore fundamental principles of molecular transport, assembly and motility under crowded cytoplasmic environments~\cite{19, 20}.

In this Letter, we demonstrate that forces generated by enzymatic reactions are sufficiently long-ranged and strong enough to influence the dynamics of their surroundings, under artificial crowded environments. This was demonstrated by using crowded colloidal suspensions of 3~$\mu$m polystyrene particles mixed with solutions of active enzymes like urease, where the amount of crowding was controlled by changing the area fractions of the suspended microparticles. At shorter time-scale regimes, the passive tracers were found to display diffusive dynamics while at longer time scales, their behavior was found to be sub-diffusive. With increased amount of crowding, the diffusion coefficient at shorter time scale and the diffusion exponent at longer time scale decreased gradually, as observed in other previous studies~\cite{21, 22}. However, with the onset of enzymatic reactions in the system (triggered by the addition of calculated amount of substrate solutions from outside), both the diffusion coefficients at shorter time scale and the diffusion exponents at longer time scale were found to increase. The recovery of diffusion values and exponents were likely due to the decrease in effective viscosity and particle caging effect respectively -- both facilitated by the enzyme substrate reactions. Experiments were also conducted with enzyme-coated microparticles, at sufficiently higher area fraction limits. Even on this occasion, substrate turnover was found to generate sufficient mechanical force to enhance the diffusivity of the particles and influence their sub-diffusive dynamics at longer time scales. We consider both these observations as significant since from a scientific standpoint, although the co-operativity between diffusing enzymes in various intracellular signalling pathways has been well documented, the degree to which their activity plays a role in cellular mechanics has not yet been investigated. Moreover, although it was hypothesized that localized energy transduction by enzymes was capable of generating long-range dynamic interactions with their surroundings, even in crowded conditions~\cite{23, 24, 25}, to date, there has been no experimental studies validating such propositions. The previous theoretical work based on Langevin dynamics of active dimers~\cite{25} seems to correspond to the experiments performed with the enzyme-coated microparticles in this study. However, to understand the experimental results obtained with passive microparticles in active enzyme suspensions, new simulations have been performed. Our results, therefore, promises a distinct shift in paradigm in molecular biophysics research, whereby localized energy transduction by enzymes is expected to play a crucial role in understanding diffusion-mediated intracellular processes. The catalysis-induced force generation and recovery of particle diffusion under artificial crowded environment may also provide newer insights into the reported stochastic motion of the cytoplasm~\cite{26, 27}, glass transition of cytoplasmic matrix during metabolism~\cite{28, 29}, and dynamics of nanoswimmers~\cite{30} and convective transport in cells~\cite{31, 32}, opening up new research opportunities in active biomolecular mechanics~\cite{33}.

%\section{\label{sec:level2}RESULTS AND DISCUSSIONS}\protect\

\begin{figure}[t]
    \centering
    \includegraphics[width=0.49\textwidth]{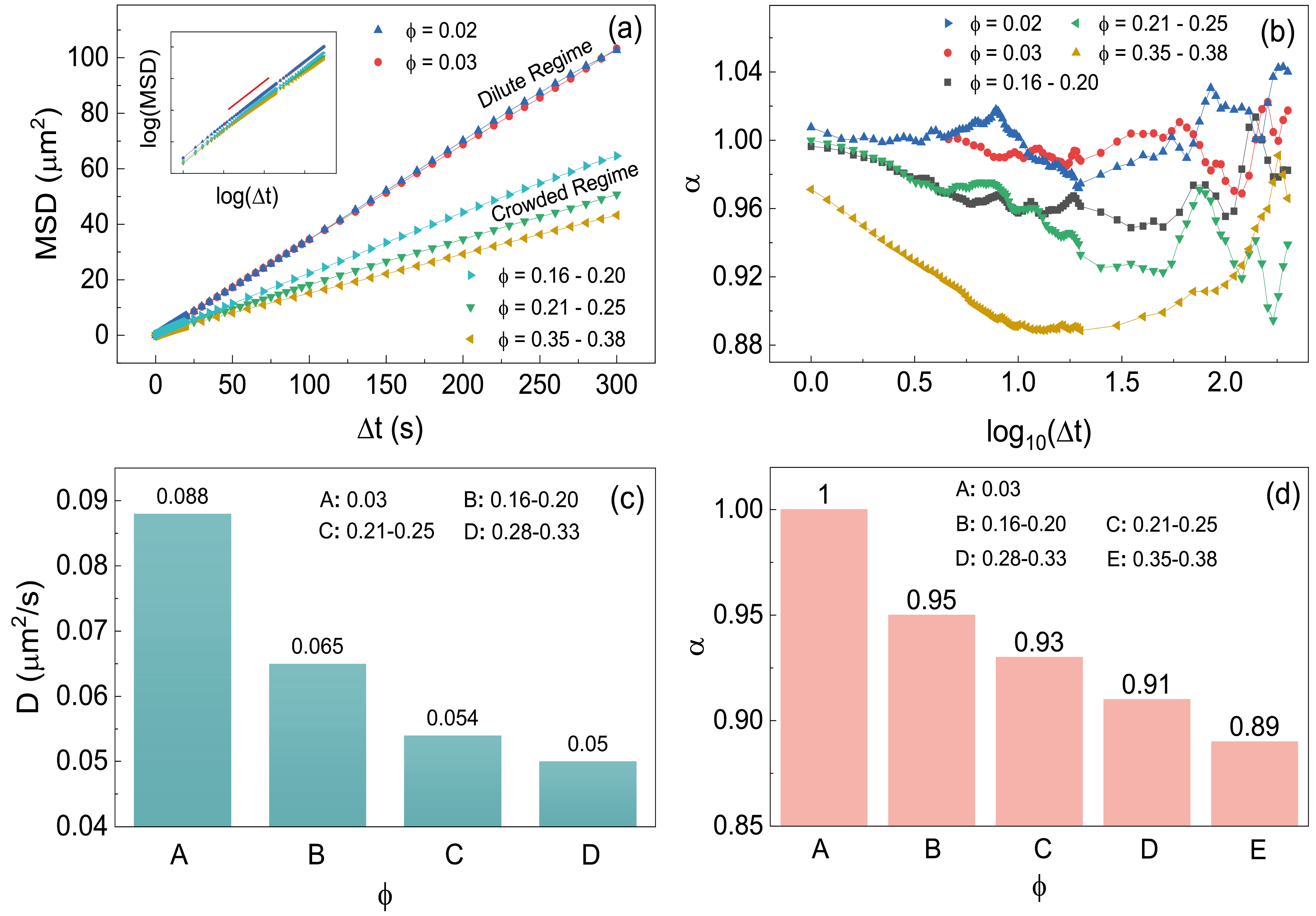}
    \caption{(a) The MSD profiles of microparticle suspensions without activity for different area fractions. The inset shows the same plots in the log scale. (b) Diffusion exponents $\alpha$ as a function of $\log(\Delta t)$, which helped in identifying the short and long time-scale regimes. The variation of short time-scale diffusion coefficient and long time scale diffusion exponent are shown in (c) and (d), respectively. Particles with $\phi$ = 0.35--0.38 did not show any diffusive motion for the entire range of the time steps used, and as such it has not been included in (c). Moreover, data corresponding to $\phi$ = 0.28--0.33 has been included in (c) and (d) only.}
    \label{fig:fig1} 
\end{figure}

We measured the mean squared displacement (MSD) of 3~$\mu$m polystyrene tracer particles in deionized water for five different area fractions $\phi$ (0.02, 0.03, 0.16--0.20, 0.21--0.25, 0.35--0.38). Maintaining stable particle area fraction values during the experiments was challenging, as the particles were in motion. To ensure that the experimental measurements were statistically significant, we tracked hundreds of particles in each data set and performed necessary control experiments, the details of which are given in the Supplementary Material (SM)~\cite{SM}. For low area fractions, particle motion remained mostly diffusive, whereas at higher values of $\phi$, the plots showed sub-diffusive behavior with gradually decreasing slopes with increasing time steps (Fig.~\ref{fig:fig1}a). Considering the MSD changes in the log scale (Fig.~\ref{fig:fig1}a, inset), we observed that for $\phi \le 0.03$, the motion of the particles remained diffusive at all time steps. For higher values of $\phi$, until a time step of 3~s, the particle motion remained diffusive. However, at intermediate time steps (20--50~s) the particle dynamics was dominated by the crowding in the system making their motion sub-diffusive in nature. Interestingly, at the highest area fraction ($\phi$=0.35--0.38) limit, the particles did not show any diffusive motion for the entire range of the time steps used. At sufficiently longer time steps, the particle motion became diffusive again. These observations were further confirmed by calculating $d \log(\mathrm{MSD})/d \log(\Delta t)$ (which yielded the corresponding diffusion exponent $\alpha$) and plotting it as a function of $\log(\Delta t)$ (Fig.~\ref{fig:fig1}b). For data analysis, we considered the time intervals $\Delta t = 1$--3~s as the short time scale regime where the motion remained diffusive. The intervals $\Delta t$ = 20--50~s was chosen as the long-time scale regime where the crowding effect dominated.

In dilute particle concentration limit, the diffusion coefficient of $3~\mu \mathrm{m}$ polystyrene particles was measured to be $0.088~\mu \mathrm{m}^2/\mathrm{s}$. However, from the Stokes-Einstein relation the diffusion coefficient was estimated to be $0.14~\mu \mathrm{m}^2/\mathrm{s}$ for the same particle at room temperature T = 25~${}^\circ$C in water (viscosity $\eta=1~\mathrm{cP}$). Therefore, the experimentally measured diffusion coefficient is 37\% less than the expected value for infinite dilution. This could be explained by considering the effect of the bottom surface and corresponding hindrance in particle diffusion~\cite{34, 35}. The negatively charged microparticles could interact with the negatively charged surface~\cite{36, 37}, leading to the restricted diffusion of the former (see details in the SM~\cite{SM}). We also estimated the increase in effective viscosity of the particle suspension given by $\eta (\phi) = \eta_0 (1+2.5\phi)$~\cite{38} and hypothesized that the observed decrease in the diffusion coefficient at the shorter time-scale regime was due to the increase in effective viscosity with higher area fraction $\phi$ (Fig.~\ref{fig:fig1}c). The supporting calculations are given in SM~\cite{SM}. We also hypothesized that the increased $\phi$ resulted in greater degree of caging effect~\cite{39, 40} at longer time-scale regimes, resulting in lowering the sub-diffusive exponent $\alpha$ with $\phi$, as observed (Fig.~\ref{fig:fig1}d). 

\begin{figure}[t]
    \centering
    \includegraphics[width=0.475\textwidth]{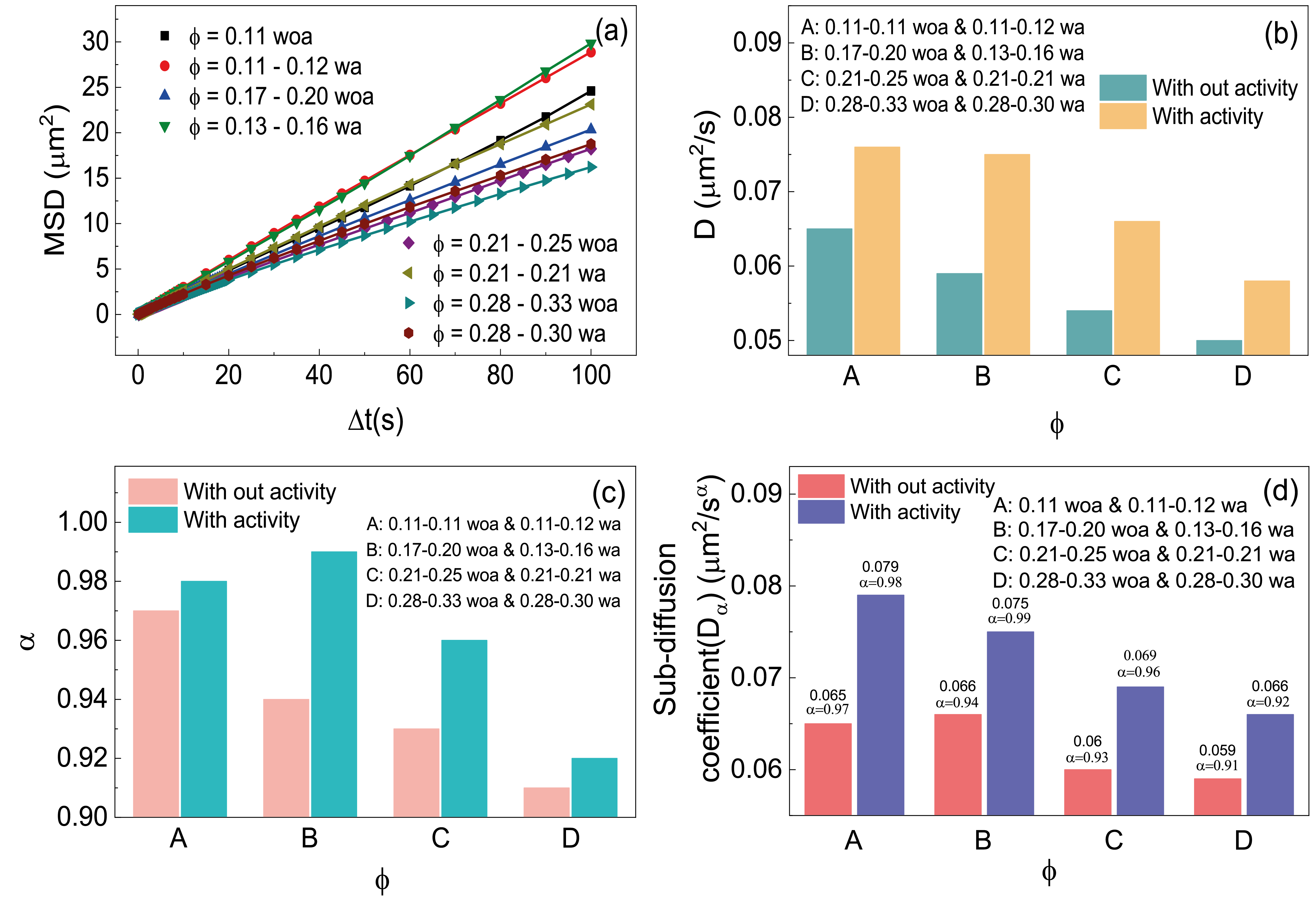}
    \caption{(a) The MSD profiles of microparticle suspensions for different area fractions in the presence (wa) and absence (woa) of free enzyme activity in the system. The corresponding short time-scale diffusion coefficients and long time-scale diffusion exponents are shown in (b) and (c), respectively. (d) Sub-diffusion coefficients $D_\alpha$ at various area fraction values.}
    \label{fig:fig2}
\end{figure}

From the above results, it became clear that crowding in a colloidal system could significantly influence both the short time diffusion coefficient and long-time diffusion exponents of suspended particles. To check if enzymatic catalysis could generate sufficient forces to counter this sub-diffusive behavior and restore the diffusive dynamics of the system under crowded conditions, we performed experiments both with tracers suspended in active enzyme solutions and high concentrations of enzyme functionalized microparticles in substrate-rich media. A recent study has demonstrated enhanced propulsion of catalase powered motor confined within giant unilamellar vesicles~\cite{41}, wherein we demonstrate in this study that similar enhancement in motion could also be realized for passive particles suspended in active crowded environment. We selected molecules of active urease as nanomotors in our system owing to their robustness, and high substrate turnover rate at room temperature (see SM)~\cite{SM}. As both ensemble and time averages were considered for MSD calculations, care was taken to fix the reaction rate that allowed the catalytic reaction to continue for a significant duration. Also, the enzyme substrate concentrations were chosen in such a manner that ensured sufficient substrate turnover and generation of nearly constant mechanical forces during the entire measurement period. Under crowding conditions, the MSD showed enhanced tracer dynamics in the presence of enzymatic activity (Fig.~\ref{fig:fig2}a).  To observe the change in the diffusive parameters in presence of substrate turnover, we measured the particle diffusion coefficients at shorter time scales (1--3~s), and diffusion exponents at longer time scales (20--50~s) and compared them with those measured in absence of catalysis. Fig.~\ref{fig:fig2}b shows the diffusion coefficients measured at different crowding conditions in the presence and absence of enzymatic activity. Force generated by free enzymes in solution was found to enhance the tracer diffusion by 15--25$\%$ which was, as mentioned earlier, likely due to the lowering of effective viscosity in the presence active enzyme propulsion. At longer time-scale regimes, the diffusion exponents of the particles showed a 1--5$\%$ enhancement, in the presence of catalytic activity (Fig.~\ref{fig:fig2}c). The corresponding sub-diffusion coefficients~\cite{42} are given in Fig.~\ref{fig:fig2}d. This indicated that in the presence of force generated due to substrate catalysis, the particles were able to get themselves freed from the crowding effects imposed by their neighbors and displayed enhanced diffusive dynamics. To confirm that the free enzymes did not adsorb over the polymer bead surface during experiments and influenced their propulsion, we also performed experiments with microparticles coated with a thin layer of bovine serum albumin (BSA), which also showed similar enhancement in particle motion during catalysis. The details of the experiments performed and results obtained are given in SM~\cite{SM}. It was also noted that upon complete depletion of the substrate in the experimental chamber, both the tracer diffusion at short time-scale regime and diffusion exponent in the long time-scale regime decreased again, like in passive crowded suspensions, indicating that the particles started feeling the effect of crowding in absence of the force generated by the enzymes. The results are given in SM~\cite{SM}.  
\begin{figure}[t]
    \centering
    \includegraphics[width=0.49\textwidth]{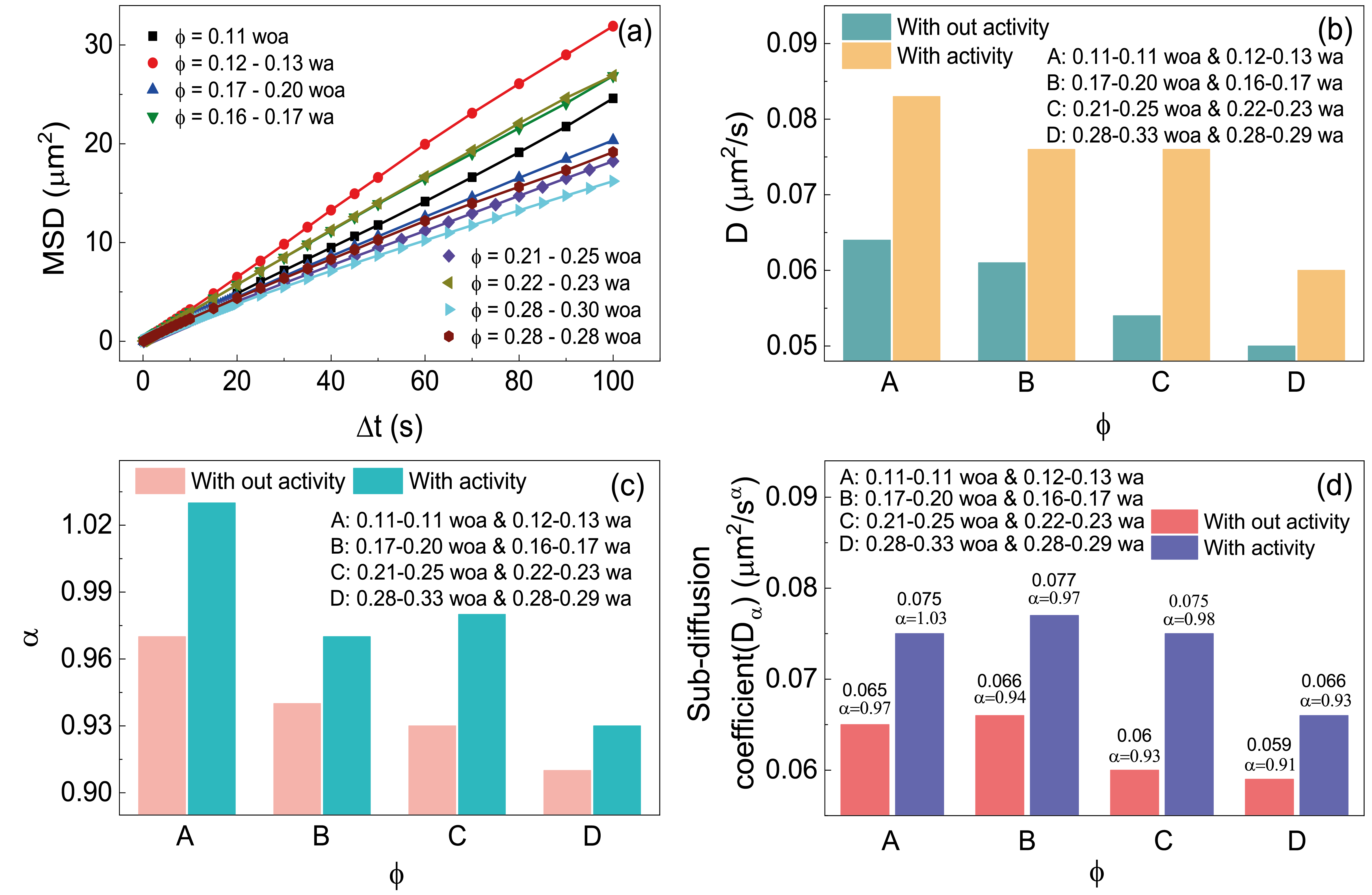}
    \caption{(a) The MSD profiles of enzyme-immobilized microparticle suspensions for different area fractions in the presence (wa) and absence (woa) of substrate solution in the system. The corresponding short time-scale diffusion coefficients and long time-scale diffusion exponents are shown in (b) and (c), respectively. (d) Sub-diffusion coefficients $D_\alpha$ at various area fraction values.}
    \label{fig:fig3}
\end{figure}
Like molecules of free enzymes, microparticles coated with immobilized active enzymes have been reported to behave as motors and display nontrivial collective dynamics~\cite{43}. Instead of using free enzyme molecules as mechanical energy sources to counter the effect of crowding, we used an assembly of urease functionalized active microparticles and investigated if during substrate turnover, the particles could generate forces to overcome their mutual crowding effects. Although theoretical analysis performed earlier suggested such a possibility,~\cite{25} to the best of our knowledge, it has not yet been demonstrated experimentally. Like passive microparticles suspended in active enzyme solution, the recovery of diffusive dynamics was also observed with active microparticles in different crowding conditions. The microparticles were functionalized with active urease enzymes using biotin streptavidin linkage chemistry (see SM for details~\cite{SM}). Diffusion studies were performed using different area fractions of particles that corresponded to different degrees of crowding. Fig.~\ref{fig:fig3}a shows the MSD plots of the particles while Figs.~\ref{fig:fig3}b and c show the short-time diffusion coefficients and long-time diffusion exponents measured at different area fractions. Clearly, like the particle in active enzyme suspensions, the enzyme functionalized particles were able to generate sufficient mechanical forces, overcome the effect of crowding to a significant degree and restore their diffusive dynamics. In case of enzyme coated particles, the short-time diffusion coefficients was found to get enhanced by 20--40$\%$ in the presence of substrate catalysis, while the long-time diffusion exponents increased by 2--6$\%$. The corresponding sub-diffusion coefficients are given in Fig.~\ref{fig:fig3}d. From the experimental results, we therefore concluded that substrate catalysis by active enzyme molecules generated sufficiently large forces that could influence the dynamics of their surroundings under artificial crowded conditions. The effective viscosity estimated at the short time-scale regime due to crowding was found to be in the range of cytoplasmic viscosity~\cite{44, 45}. The enhanced diffusion of particles observed at this time-scale regime indicated that the force generated due to catalytic turnover were able to lower the effective viscosity, thereby enhancing the particle propulsion. 
%\section{\label{sec:level3}NUMERICAL STUDIES}\protect\

\begin{figure}[h]
    \centering
    \includegraphics[width=0.49\textwidth]{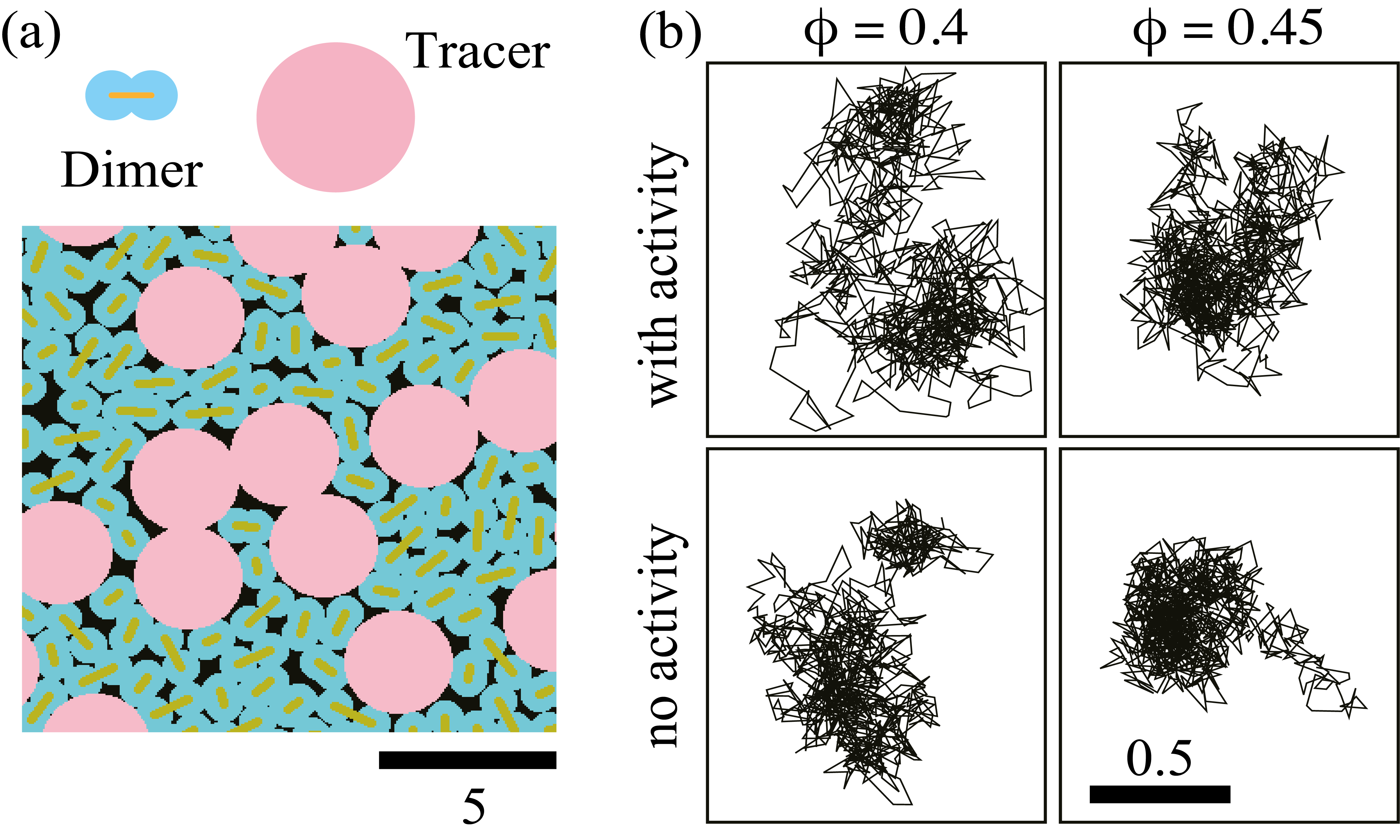}
    \caption{(a) Snapshot of the suspension composed of the tracer (pink particles) and the dimers (light blue particles connected by yellow bonds). (b) Trajectories of the tracer particles. A spread (compact) trajectory are observed in the case for the lower (higher) area fraction with (without) activity, which reflects the magnitude of caging effect. The spatial scales for 4 panels are common.}
    \label{fig_num_1}
\end{figure}

To understand the diffusion enhancement of passive tracers by the active enzymes as shown in Fig.~\ref{fig:fig2} from the microscopic viewpoint, we considered a two-dimensional model composed of tracers surrounded by dumbbell-shaped particles (Fig.~\ref{fig_num_1}a).
The dumbbell-shaped particles changed their arm length incoherently, which roughly imitated the conformation changes in enzymes during catalysis.
Hereafter, the dumbbell-shaped particles are called the dimers.
The mathematical formulation corresponding to the tracer dynamics in active dimer suspension is as follows.
In our model, the Langevin dynamics of the center positions of the tracer particles $\bm{R}_i$ and the beads consisting the dimer $\bm{r}_j^{(n)}$ with excluded volume effect are considered, where $i (= 1, \cdots, M)$, $j (= 1, \cdots, N)$ and $n (= 1,2)$ indicate the indices for a tracer, dimer, and bead consisting the dimer, respectively.
The dynamics of the tracer particles are governed by the following over-damped Langevin equation:
\begin{align}
\frac{d \bm{R}_i}{dt} = -\mu_t \frac{\partial U}{\partial\bm{R}_i} + \bm{\xi}_{t,i}(t) \label{eq_tracer}
\end{align}
where $\mu_t$ is the mobility. The term $\bm{\xi}_{t,i}$ is the thermal noise which satisfies $\left< \xi_{t,i,\alpha} (t) \right> = 0$ and $\left< \xi_{t, i,\alpha} (t) \xi_{t,j,\beta} (s) \right> = 2 \mu_t k_B T \delta_{ij} \delta_{\alpha\beta} \delta(t-s)$ $(\alpha, \beta = x,y)$.
The function $U$ denotes the potential reflecting the excluded volume effect of particles
\begin{align}
U =& \frac{1}{2} \sum_{i=1}^{N} \sum_{j (\neq i) = 1}^{N} \sum_{m=1}^{2} \sum_{n=1}^{2} u \left( \left|\bm{r}_i^{(m)} - \bm{r}_j^{(n)} \right|; 2 r_0 \right) \nonumber \\
&+ \frac{1}{2} \sum_{i=1}^{M} \sum_{j (\neq i) = 1}^{M} u \left( \left|\bm{R}_i - \bm{R}_j \right|; 2 R_0 \right) \nonumber \\
&+ \sum_{i=1}^{N} \sum_{j=1}^{M} \sum_{m=1}^{2} u \left( \left|\bm{r}_i^{(m)} - \bm{R}_j \right|; r_0 + R_0 \right) 
\end{align}
where
\begin{align}
u(r; \rho_0) = \left \{
\begin{array}{ll}
u_0 (\rho_0 - r)^2, & (r < \rho_0) \\
0, & (r > \rho_0)
\end{array}
\right.
\end{align}
Here, $R_0$ and $r_0$ are the radii of the tracer and the bead consisting dimers.
Thus, the first term in the right side of Eq.~\eqref{eq_tracer} represents the repulsive force during particle collision.
In the same way, the dynamics of the dimer are given by the following over-damped Langevin equation:
\begin{align}
    \frac{d \bm{r}_i^{(n)}}{dt} = -\mu \frac{\partial E_i}{\partial \bm{r}_i^{(n)}}-\mu \frac{\partial U}{\partial\bm{r}_i^{(n)}}+ \bm{\xi}_i^{(n)}(t) \label{eq_dimer}
\end{align}
where $\mu$ is the mobility.
The term $\bm{\xi}_{i}^{(n)}$ is the thermal noise, which satisfies $\left< \xi_{i,\alpha}^{(m)} (t) \right> = 0$, $\left< \xi_{i,\alpha}^{(m)} (t) \xi_{j,\beta}^{(n)} (s) \right> = 2 \mu k_B T \delta_{mn} \delta_{ij} \delta_{\alpha \beta}\delta(t-s)$.
It should be noted that the particle mobilities hold the relation $\mu = R_0 \mu_t / r_0$ from the Stokes' law.
The first term in the right side of Eq.~\eqref{eq_dimer} expresses the force between the beads composing a dimer, where
\begin{align}
    E_i(t) = \frac{k}{2} \left (\left | \bm{r}_{i}^{(1)} - \bm{r}_{i}^{(2)} \right | - \ell_i(t) \right )^2
\end{align}

In the numerical calculations, we compared two cases: one is that the length of the dimer, $\ell_i(t)$, changes in time, which corresponds to the conformation changes in the enzymes. Specifically, it changes as follows
\begin{align}
    \ell_i(t) = \ell_0 + \ell_1 \sin \psi_i(t)
    \end{align}
    \begin{align}
    \frac{d \psi_i}{dt} = \omega + \zeta_i(t)
\end{align}
$\zeta_i(t)$ is the white Gaussian noise with $\left< \zeta_i(t) \zeta_j(s) \right> = 2 \eta \delta_{ij} \delta(t- s)$.
The other case corresponds to the tracer diffusion in the absence of enzymatic catalysis, i.e., $\ell_i$ has a constant value $\ell_0$.
We also investigated the dependence of diffusivity on the particle density to understand the crowding effect on the tracer particles.

The spatial scale is normalized by the bead radius comprising a dimer, $r_0$.
The radius of the tracer particle is $R_0 = 3$, and the the dimer's natural length is $\ell_0 = 1.5$.
The time is scaled by the damping of the bead consisting of a dimer, i.e., $\mu = 1$.
Other parameters are set to be $\ell_1 = 1.0$, $u_0 = 1$, $k=1$, $\omega = 0.1$, $k_B T = 0.01$, and $\eta = 0.1$. The area fraction of dimer was fixed to be 0.5, and that of tracer particles was changed as 0.4, 0.4125, 0.425, 0.4375, and 0.45.

\begin{figure}[t]
    \centering
    \includegraphics[width=0.49\textwidth]{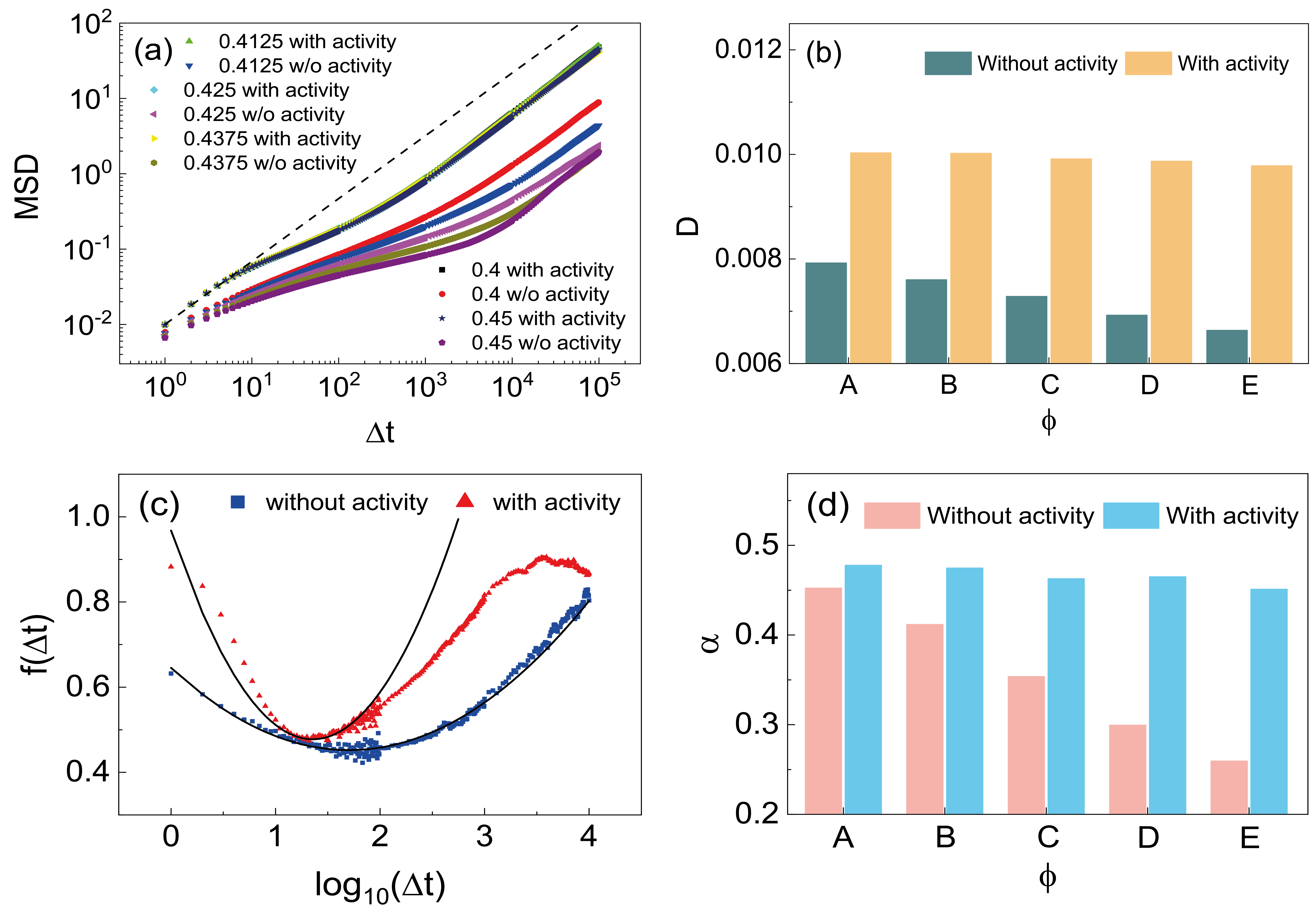}
    \caption{(a) log-log plot of MSD against time interval $\Delta t$. The black dashed line shows the theoretical MSD for a single tracer particle without collisions. The plots in the case with activity are almost collapsed. (b) Dependence of diffusion constant $D$ on the area fraction and activity. (c) The derivative of $\log(\mathrm{MSD})$ plotted against $\log(\Delta t)$. The cases with the area fraction $\phi = 0.4$ are exemplified. (d) Sub-diffusion exponent $\alpha$ depending on the area fraction $\phi$. In (b) and (d) the area fractions are A:0.4, B:0.4125, C:0.425, D:0.4375, E:0.45.}
    \label{fig5}
\end{figure}

By increasing the area fraction, the trajectory of the tracer particle becomes compact, while the activity makes the trajectory broader, as shown in Fig.~\ref{fig_num_1}b. To elucidate this further, the MSDs of the tracer particles as  function of time intervals
were obtained by averaging over all tracer particles. As seen in Fig.~\ref{fig5}a, the MSDs become smaller for larger area fractions of tracer particles in the absence of any dimer activity. The diffusion constant determined by the MSD at $t = 1$ is shown in Fig.~\ref{fig5}b.

Since the diffusion constant should coincide $D_0 = \mu_t k_B = 10^{-3}/3$ for the limit of $t \to 0$, the crowding effect already appears at $t = 1$.
To check the sub-diffusion regime, the local gradient of MSD, $f(\Delta t) = d[\log(\mathrm{MSD}(\Delta t))]/d(\Delta t)$ was checked.
The examples of $f(\Delta t)$ are plotted in Fig.~\ref{fig5}c, which are qualitatively similar to the experimental results in Fig.~\ref{fig:fig1}b.
We can see that the variations seem to take minimum values around $\Delta t \simeq 1.3$ (1.8) in the case with (without) activity.
Since $f(\Delta t)$ is noisy, the sub-diffusion exponent was defined as the local minimum of the quadratic-function fitting of $f(\Delta t)$.
The fitting range was $\Delta t_\mathrm{min}-0.5 \leq \Delta t \leq \Delta t_\mathrm{min}+0.5$, where $\Delta t_\mathrm{min}$ gives the minimum value of raw $f(\Delta t)$.
As shown in Fig.~\ref{fig5}d, the sub-diffusion exponent $\alpha$ estimated in the presence of activity always exceeds that estimated without dimer activity. Moreover, the sub-diffusion exponent is a decreasing function of the area fraction of tracer particles for both active and non-active cases.
Such trend is qualitatively the same as the experimental results.
From the numerical results, we assert that the conformation changes in the enzymes is essential in deciding the dynamics of tracer particles in active crowded suspensions.

%\section{\label{sec:level4}CONCLUSIONS}\protect\
In summary, we demonstrate that the force generated by enzymes during substrate turnover is sufficiently strong to influence the dynamics of their surroundings, under artificially crowded environments. These observations have several important implications and offer opportunities to investigate the consequences of biomolecular activities over intracellular transport, assembly and organization of components under crowded cytosolic environments.

\begin{acknowledgments}
KKD thanks SERB, India (ECR/2017/002649), DST, India (DST/ICD/BRICS/PilotCall3/BioTheraBubble/2019) and IIT Gandhinagar for financial supports. HK was supported by JSPS KAKENHI Grant Number JP21H01004. We are thankful to Prof. Alexander Mikhailov and Prof. Raymond Kapral for insightful discussions. Help received from Dr. Chandan Kumar Mishra in developing the particle tracking methodology is gratefully acknowledged.  
\end{acknowledgments}

\bibliography{bibfile}

\end{document}